\documentclass[conference]{IEEEtran}
\makeatletter
\def\ps@headings{%
\def\@oddhead{\mbox{}\scriptsize\rightmark \hfil \thepage}%
\def\@evenhead{\scriptsize\thepage \hfil \leftmark\mbox{}}%
\def\@oddfoot{}%
\def\@evenfoot{}}
\makeatother
\usepackage{amsmath,amsfonts}
\usepackage{algorithm}
\usepackage{array}
\usepackage{textcomp}
\usepackage{stfloats}
\usepackage{url}
\usepackage{verbatim}
\usepackage{graphicx}
\usepackage{cite}
\usepackage{algpseudocode} 
\usepackage{caption}
\usepackage{color}
\usepackage{subcaption}
\usepackage{lipsum}
\usepackage{geometry}

\begin{document}
\title{
%QoS-Driven Adaptive and Resilient UAV Base Station Formations\\
\vspace{-0.0in}
%QoS Driven Proactive and Resilient Orchestration of UAVs Differentiated Services 
Proactive and Resilient UAV Orchestration for QoS Driven Connectivity and Coverage of Ground Users
%Enabled Wireless Networks
%Dynamic UAV Orchestration for On-Demand Connectivity with QoS Guarantees%Differentiated Service Levels
}

\author{Yuhui~Wang and Junaid~Farooq\\
Department of Electrical and Computer Engineering, \\
University of Michigan-Dearborn, Dearborn, MI 48128 USA, \\
{Emails: \{ywangdq, mjfarooq\}@umich.edu}.
\vspace{-0.0in}
%\thanks{The authors are with the Department of Electrical and Computer Engineering,
%University of Michigan-Dearborn, Dearborn, MI 48128 USA (emails: ywangdq@umich.edu; mjfarooq@umich.edu). }
}

\maketitle

\begin{abstract}
Unmanned aerial vehicles (UAVs) are being successfully used to deliver communication services in applications such as extending the coverage of 5G cellular networks in remote areas, emergency situations, and enhancing the service quality in regions of dense user populations. While optimized placement solutions have been proposed in literature for ensuring quality of service (QoS) of users, they may not be ideal in highly mobile, autonomous, and diverse network scenarios. This paper proposes a proactive and resilient framework for distributed and dynamic orchestration of UAV small cells to provide QoS differentiation in the network. The UAV locations are tailored to end-user locations and service needs while ensuring that UAVs maintain localized backhaul connectivity. Simulation experiments show that under scenarios where physical placement can achieve service differentiation, the developed framework leads to a stable configurations of UAVs satisfying above 90\% of user QoS requirements.
\end{abstract}

\begin{IEEEkeywords}
Unmanned Aerial Vehicles, Connectivity, Resilience, Quality-of-Service, Swarming.
\end{IEEEkeywords}

\IEEEpeerreviewmaketitle

\vspace{-0.0in}
\section{Introduction}
\begin{figure}[t]
    \centering
    \vspace{-0.0in}
    \includegraphics[trim={0 0 0 0 in},clip,width=2.5in]{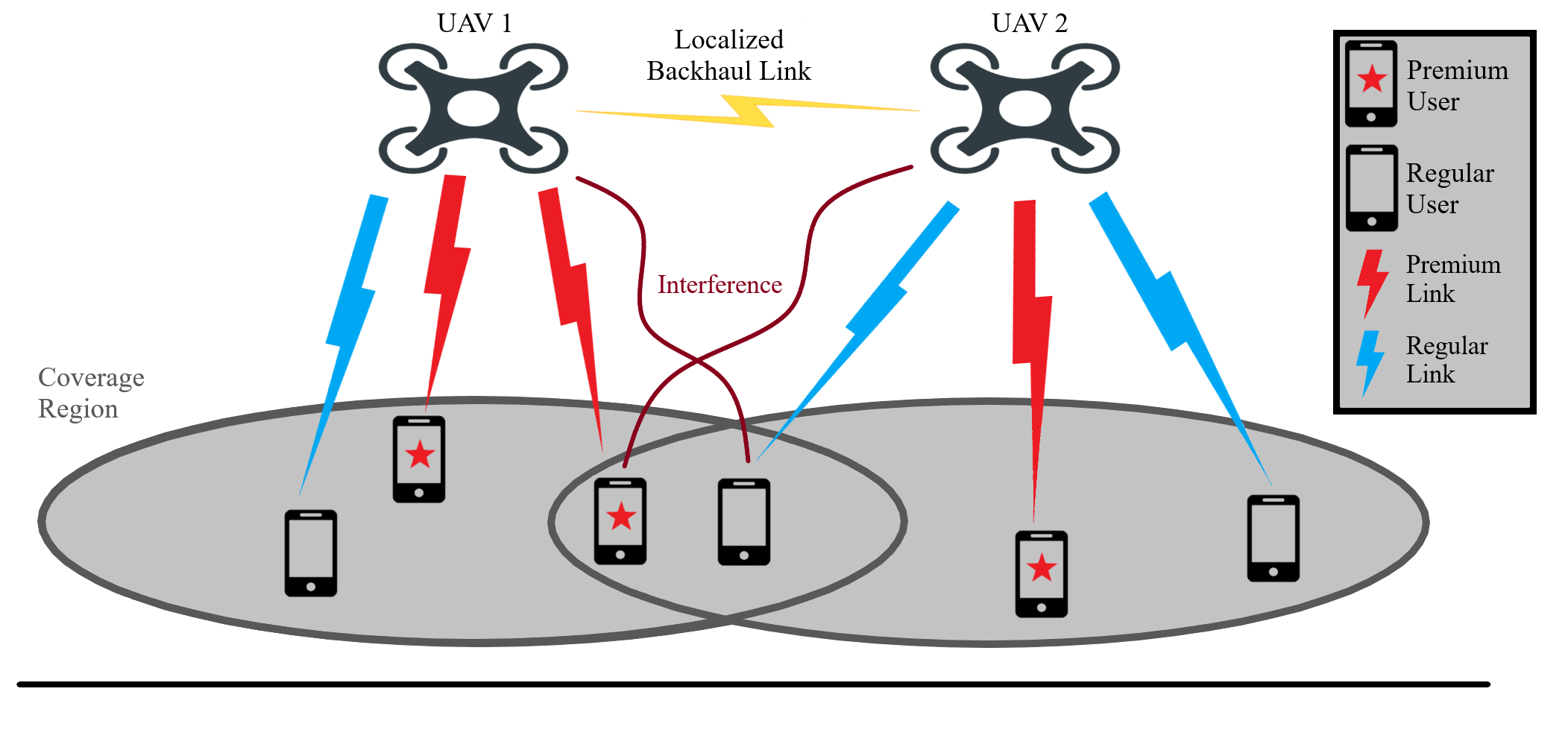}
    \caption{Orchestration of UAVs in the air to serve premium and regular ground users with different QoS requirements.\vspace{-0.2in}}
    \label{fig:network}
\end{figure}

UAVs have proven to be a convenient choice for providing on-demand wireless connectivity in disasters and emergency scenarios~\cite{yuhui_icc}. They are also critical in enhancing and improving the service gaps in fifth generation (5G) cellular networks. In certain situations such as public events, gatherings, parades, or sporting events, there may be a need to supplement the cellular networks with small cell base stations (BSs) mounted on UAVs~\cite{tech_report_uas, placement_QoS_guaranteed}. While several factors play a role in the quality of communication experienced by the users connected to the UAVs, the placement the UAV BSs can be crucial in ensuring the strength of wireless signals between the UAV and the ground users~\cite{UAV_tutorial}. Moreover, multiple types of users may be present in the network, e.g., those that have subscribed for higher speeds and preferred resource allocation, hereby referred to as \emph{premium users}, and those that are using standard service, hereby referred to as \emph{regular users}. Therefore, the placement of UAVs needs to be optimized to cater for the quality of service (QoS) requirements of different users~\cite{3D_QoS,QoS_provisioning}. An illustration of the considered network scenario is shown in Fig.~\ref{fig:network}, where the premium user and regular user are connected to a network of UAV BSs that are orchestrated in the air to meet the data needs of the users.

%The demand for 5G services  scenarios can be different, e.g., in parades, games, and congested areas, 
In conventional cellular networks, the BS locations are fixed and there is not much that can be done to improve the link quality of a particular user from a path-loss or interference standpoint. However, in the case of a mobile aerial BS, there is a high degree of freedom in the location of the BS, allowing it to move closer to a particular user and vice versa to satisfy QoS requirements~\cite{user-centric-deployment}. For instance, if the user has a premium subscription and if it does not achieve sufficient data rates due to interference or other channel impairments, the network can reconfigure itself to improve the QoS of the user. In essence, we can leverage the mobility and flexibility of aerial placement in UAVs to provide a customized service to ground users. This enables new service models and pricing options by the network operators that will trigger higher revenue generation. Existing cellular operators can also extend their network by adding on-demand small cells in scenarios where a higher user density is expected~\cite{On-demand_small_cells}. 

There are several existing works in the literature that consider the QoS provisioning in the placement and resource allocation decisions of UAVs~\cite{3d_location_resource_allocation_optimization,Delay_and_error_QoS,Backhaul_User_QoS_constraints,particle_swarm_optimization,edge_prior,energy_efficient_1,QoS_guarantee_1}. These consider factors such as energy efficiency~\cite{energy_efficient_1,3d_placement}, QoS guarantees~\cite{QoS_guarantee_1,3d_location_resource_allocation_optimization,Delay_and_error_QoS}, and trajectory optimization~\cite{Backhaul_User_QoS_constraints,RL,particle_swarm_optimization}.
However, there are several limitations: (i) The existing solutions are centralized and are not based on feedback from the users, whereas our proposed framework proactively involves the performance of users into the placement decisions via real-time feedback on the achieved data rates. (ii) They are not adaptive to changes in user locations and performance needs over time, whereas our proposed solution is dynamic and continuously adapts to the highly mobile, autonomous, and dynamic network scenarios. Hence it is resilient to potential failures and cyber-physical attacks on the network. (iii) Existing works focus on the placement decisions while ignoring the details of UAV dynamics and control aspects, whereas we have used a dynamical model to realistically mimic the practical scenarios and used a controller based approach inspired from swarming and flocking of robotic systems.
%Solutions based on numerical optimization techniques such as~\cite{particle_swarm_optimization} are not practical due to convergence issues over dynamic networks. 

%open-loop UAV placement strategy without taking the real-time QoS of users into account. We need a feedback oriented framework that can adapt the configuration of aerial BS according to the needs of the ground users. 
 
Although the problem of optimized placement of UAVs can be solved in a centralized manner, it becomes less applicable in complex real world scenarios where a centralized planner may not be available to have a global view of the network. We propose a completely distributed and dynamic approach to orchestrate the UAVs using an incentive mechanism design approach that facilitates the fulfilment of QoS requirements of the users. 
%. No centralized planning is needed and the UAV network is self-configurable, flexible and adaptive.
%We propose an incentive mechanism design that results in a positioning of UAVs satisfying the data rate requirements of the users. It is tailored towards the needs of the users.  
%If the users are mobile, e.g., in a parades or marches, the aerial formation can adapt itself to ensure the required service delivery.
%- 5G and 6G communication needs and increasing demand for data.
%- Differentiated service requirements of users
%- Small cells and heterogeneous networks are one way to improve data rates
%- Physical-layer performance is dependent on the location of UAVs and resulting distance to the users. This is increasingly more important with the emerging mmWave communication for ultra-high data rates. 
%- UAV is a convenient mobile platform to provide on-demand connectivity. Already tested and utilized in emergency situations.
%the cellular performance is low and it is impractical to have ubiquitous deployment of small cells.
%We propose an iterative algorithm to achieve this objective. Traditionally, centralized optimization approaches are used.
%It adapts according to user performance levels and the target performance. 
%user requirements can change over time. Aso the location of users is mobile. 
%QoS provisioning is becoming an area of importance in UAV communication networks.
Moreover, our framework can also be augmented to centralized placement solutions to further fine tune the performance of users in the network and adapt to their changing positions and QoS needs. 
%It can be used by 5G operators when deploying UAV-based small cells to supplement traditional cellular coverage.
The key contributions of this work are as follows:
\begin{itemize}
    \item We develop a proactive and resilient dynamic framework for real-time UAV BS orchestration that provides QoS differentiation for premium and regular ground users. 
    \item We propose a UAV swarming approach that ensures localized backhaul connectivity among UAVs while adaptively orienting themselves to cater for user QoS requirements. 
    %then formulate the problem of UAV positioning to satisfy QoS requirements of users while maintaining localized backhaul as a capacitated facility location problem. Since the problem is NP-hard, we propose a low complexity, distributed approach inspired from swarming and flocking behaviors in nature to solve the problem. 
    \item We conduct numerical experiments in several simulation scenarios to test our solution. The results demonstrate above 90\% efficacy of the proposed algorithm. 
    %in scenarios where a physical orchestration can satisfy QoS requirements.
    %results depict that the
\end{itemize}

%3D deployment of BS is being done to cater for QoS needs of users.

%location and resource allocation optimization for QoS needs.

%Distributed coverage and connectivity required in various practical scenarios, e.g., emergency communication networks, battlefield communications, etc. No traditional backhaul in such situations so UAV network needs to ensure both coverage of ground users and the inter-connectivity between them.
%Furthermore, users can be arbitrarily located on the ground resulting in an additional challenge where the UAVs not only have to remain connected with each other but also need to ensure that the ground users are adequately covered.

%\section{Related Work} \label{Sec:Related_work}

The rest of the paper is organized as follows: Section~\ref{Sec:Sys_Model} provides an overview of the system model and defines the QoS provisioning problem, Section~\ref{Sec:Methodology} details the methodology of our proposed approach, Section~\ref{Sec:Results} provides simulation based validation of our framework, followed by the conclusion in Section~\ref{Sec:Conclusion}.

\vspace{-0.0in}
\section{System Model} \label{Sec:Sys_Model}
%\textcolor{blue}{Mention symbol for achieved data rate and target data rate}.
%\textcolor{red}{First talk about channel models, signal, interference, rate, etc. Then add details of dynamics and controller.}
%\textcolor{red}{UAV height? symbol h is used for both height and goal function}
Consider a set of ground users $\mathcal{M} = \{1, \ldots, M\}$ and a set of UAVs $\mathcal{U} = \{1, \ldots, U \}$, equipped with small cell BSs, to provide wireless connectivity to the ground users. The Cartesian coordinates of ground users at time $t$ are denoted by $\boldsymbol{y}(t) = [y_1(t),y_2(t),\ldots,y_M(t)]^T$, where $y_i(t) \in \mathbb{R}^3, \forall i \in \mathcal{M}, t \geq 0$. Note that the height of all ground users is assumed to be zero. Similarly, the Cartesian coordinates of the UAVs at time $t$ are denoted by $\boldsymbol{q}(t)=[q_1(t),q_2(t),...,q_U(t)]^T$, where $q_i(t) \in \mathbb{R}^3, \forall i \in \mathcal{U},t \geq 0$. The height of all the UAVs is assumed to be $H$\footnote{The altitudes of UAVs is generally regulated based on the flying environment and safety standards. Therefore, we assume a predefined constant elevation for all UAVs.}.
The velocity of the UAVs at time $t$ is denoted by $\boldsymbol{p}(t)=[p_1(t),p_2(t),...,p_U(t)]^T$, where $p_i(t) \in \mathbb{R}^3, \forall i \in \mathcal{U},t \geq 0$. Since the UAVs maintain a fixed height $H$, the vertical velocity component of all UAVs is zero.
The UAVs have a maximum communication range of $r \in \mathbb{R}^+$ such that any two UAVs can communicate only if the Euclidean distance between them is less than $r$. The neighboring UAVs of a typical UAV denoted by $i$ is represented by the set $N_i = \{j \in \mathcal{U} \backslash i : \|q_i - q_j\| \leq r \},\forall i \in \mathcal{U}$. User $m$ connects to UAV $i$ if $i=\underset{k}{\arg\min} \|q_k-y_m\|, \forall k \in \mathcal{U}$. The set of users served by UAV $i$ is represented by $\mathcal{M}_i^C$, which is limited by the maximum serving capacity $|\mathcal{M}_i^C|\leq N^{\max}$.

Each UAV uses a wireless channel $\mathcal{L}_{i},i\in\mathcal{U}$ and $\mathcal{L}_{i}\in \{0,1,2,3,...\}$ to communicate with ground users. We consider two types of users in the network denoted by $\chi$, where $\chi(m)=\text{(prem)}$ if user $m$ is a premium user and $\chi(m)=\text{(reg)}$ if user $m$ is a regular user.
%For regular users, $\chi(m)=(\text{reg})$ representing 
Regular users require a data rate $\rho^{(\text{reg})}$ while premium users require a data rate $\rho^{(\text{prem})}$. Note that $\rho^{(\text{prem})} > \rho^{(\text{reg})}$ signifying a higher data rate requirement for premium users. Regular users can only connect to UAVs through the default channel $L_0$, while premium users can access all channels $L_i$, $i \geq 0$. In the default configuration, all UAVs use channel $L_0$ to communicate with users, i.e., $\mathcal{L}_i = L_0, \forall i\in \mathcal{U}$. If a premium user cannot achieve the target data rates, the connected UAV may switch channel to reduce interference from other UAVs and consequently improve performance.

%Regular users have low target data rates $\rho^{(reg)}$ and premium users have high target data rates $\rho^{(prem)}$. 

%\subsection{UAV Dynamics Model}

\vspace{-0.0in}
\subsection{Path-Loss Model}

\begin{figure}[t]
    \centering
    \includegraphics[trim={12 0 10 0 in},clip,width=3in]{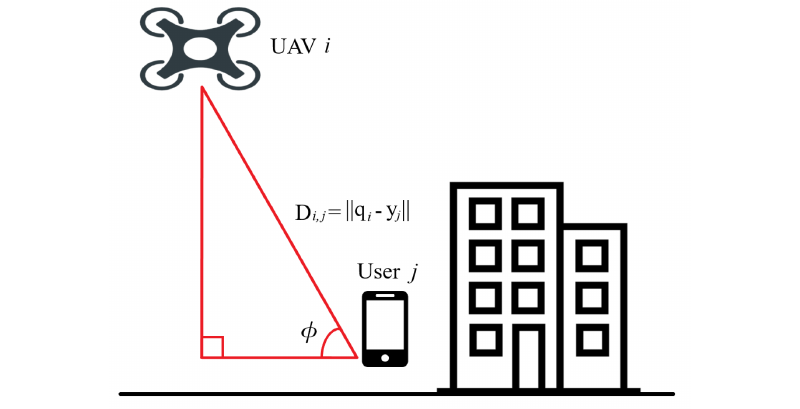}
    \caption{Relative positions of UAV and ground user. \vspace{-0.2in}}
    \label{fig:notation}
\end{figure}

The wireless signal from the UAV to the ground users may encounter a line-of-sight (LoS) path or a non-line-of-sight (NLoS) path due to reflections from buildings and other obstacles. The signal attenuation in each case is different and we use a probabilistic approach to model the availability of these paths. The probability of having a LoS path between the user and UAV is denoted by $p_{\text{LoS}}$ and the probability of having an NLoS path is denoted by $p_{\text{NLoS}}$ with $p_{\text{NLoS}}=1-p_{\text{LoS}}$. The probability of LoS connection is computed as \cite{air-to-ground-pl}:
\begin{equation}
    p_{\text{LoS}}=\frac{1}{1+\vartheta\exp\left(-\xi\frac{180}{\pi}\phi-\vartheta\right)},
\end{equation}
where $\vartheta$ and $\xi$ are constants depending on the environment and $\phi$ is the elevation angle as shown in Fig.~\ref{fig:notation}. 
%Both line of sight ()
%In scenarios that UAVs establish communications with ground users, the connections between them are steady if there are no obstructions, i.e. in line of sight (LoS). The signals may go through attenuations and reflections if there are buildings in the path, i.e. in non line of sight (NLoS). 
%The probability of the two cases are described by $\mathbb{P}({\text{LoS}})$ and 
NLoS paths may experience higher path loss as compared to LoS connections. We use the widely accepted air-to-ground path-loss model in \cite{air-to-ground-pl} to characterize the path-loss between UAVs and ground users. The path loss between a UAV $i$ and user $m$ can be expressed as:
\begin{equation}
\begin{aligned}
    PL_{i,m}=&10\log_{10}\left(\frac{4\pi f_c D_{i,m}}{c} \right)^\delta +p_{\text{LoS}}\hspace{0.03in}\eta_{\text{LoS}} +
    p_{\text{NLoS}}\hspace{0.03in}\eta_{\text{NLoS}},
\end{aligned}
\label{eq:pl_usr}
\end{equation}
where $D_{i,m}$ is the distance between the UAV and ground user, i.e.,  $D_{i,m}=\|q_{i}-y_{m}\|$, $i\in\mathcal{U}, m\in\mathcal{M}$, $c$ denotes the speed of light, $f_c$ is the carrier frequency, and $\delta$ is the path-loss exponent. The first term in \eqref{eq:pl_usr} represents the free-space path-loss, while $\eta_{\text{LoS}}$ and $\eta_{\text{NLoS}}$ are the additional average losses for LoS and NLoS paths respectively.
%which depends on the carrier frequency $f_c$, 
%3D distance between UAV and ground user $D_{i,j}=\|q_{i}-y_{j}\|$, $i\in\mathcal{U}, j\in\mathcal{M}$, 
%speed of light $c$, and path-loss exponent $\delta$. 

% Unlike the UAV to ground user link which may experience NLoS propagation, the backhaul links between UAVs are at relatively high altitudes and can be simplified as propagation in free space. Thus, we use a free space path loss model \cite{FSPL} to describe the backhaul link quality between UAV $i$ and $j$:
% \begin{equation}
%     \text{FSPL}_{i,j}=20\log_{10}\left(\frac{4\pi f_a D_{i,j}}{c}\right),
% \end{equation}
% where $f_a$ is the aerial channel frequency, $D_{i,j}=\|q_{i}-q_{j}\|$, $i,j\in\mathcal{U}, i\neq j$ is the 3D distance between UAVs. The link quality is considered good if $\text{FSPL}_{i,j}<FSPL_{max}$. As $f_a$ is fixed, the maximum communication range of UAVs $r$ is:
% \begin{equation}
%   r=\frac{c}{4\pi f_a}10^{FSPL_{max}/20}.
% \end{equation}

\vspace{-0.0in}
\subsection{Quality of Service (QoS)}
To quantify the QoS of ground users, we use the Shannon capacity based on the signal-to-interference-plus-noise ratio (SINR). The downlink SINR of UAV $i$ connecting to user $m$ can be expressed as:
\vspace{-0.0in}
%Suppose user $i$ is connected to UAV $j$. The QoS of user $i$ is measured in terms of the signal-to-interference-plus-noise ratio $\text{SINR}_{i,j}$ and 
\begin{equation}
\begin{aligned}
    \text{SINR}_{i,m}&=\frac{P_r^{i,m}}{\sigma^2+{\sum_{k \in \mathcal{U} \backslash i,\mathcal{L}_i=\mathcal{L}_k}P_r^{k,m}}},
\end{aligned}
\label{eq:sinr}
\end{equation}
with $P_r^{i,m}=P_t^{i,m}/10^{PL_{i,m}/10}$. $P_r^{i,m}$ is the received signal power by the user and $P_t^{i,m}$ is the transmission power of the UAVs. $\sum_{k \in \mathcal{U} \backslash i,\mathcal{L}_i=\mathcal{L}_k}P_r^{k,m}$ is interference power from other UAVs using the same channel $\mathcal{L}_i=\mathcal{L}_k$, and $\sigma^2$ is the thermal noise power. $P_r^{i,m},P_t^{i,m},P_r^{k,m}$ and $\sigma^2$ are all in the units of mW. Consequently, the achieved data rate of UAV $i$ connecting to user $m$, denoted by $\mathcal{C}_{i,m}$ can be written as follows:
\begin{equation}
    \mathcal{C}_{i,m}=B\hspace{0.02 in}\log_2(1+\text{SINR}_{i,m}),
\label{eq:data rates}
\end{equation}
where $B$ is the available channel bandwidth. Since user $m$ is always connected by its closest UAV, we subsequently drop the subscript $i$ for brevity of notation and use $\mathcal{C}_{m}$ to denote the QoS achieved by user $m$ while served by its associated UAV. 

\vspace{-0.0in}
\subsection{Problem Formulation}

The objective of the proposed framework is to determine a set of positions at each time instant that minimizes the difference between achieved QoS, $\mathcal{C}_m$ and target QoS, $\rho^{\chi(m)}$ for all users $m\in \mathcal{M}$, while ensuring that the spatial interval of UAVs to be between the minimum safe distance $d$ and the maximum communication range $r$. This optimization problem can be formulated as follows:
\vspace{-0.0in}
\begin{equation}
    \mathbf{P0}:\hspace{0.4 in}\min_{\textbf{q}(t)}\sum_{m\in \mathcal{M}}|\mathcal{C}_m-\rho^{\chi(m)}|,
\end{equation}
\begin{equation}
    \begin{aligned}
        \text{s.t.}\hspace{0.2 in}&\|q_i(t)-y_j(t)\|\leq r,\hspace{0.1 in} \forall i\in\mathcal{U},j\in\mathcal{M}_i^C, \\
        &d\leq \|q_i(t)-q_j(t)\|\leq r,\hspace{0.1 in} \forall i\in\mathcal{U},j\in N_i. \label{const_eq}
    \end{aligned}
\end{equation}

In problem $\mathbf{P0}$, we have a set of facilities (UAV $\mathcal{U}$) and a set of clients (users $\mathcal{M}$). Each user has a demand of target data rate $\rho^{\chi(m)}$ and is served by one UAV.
The constraints in \eqref{const_eq} signify that the Euclidean distance between any pair of UAV and its connected user, i.e.,  $\|q_i(t)-y_j(t)\|$, should be less than the maximum communication range $r$ of the UAVs. In addition, the Euclidean distance between UAVs, i.e.,  $\|q_i(t)-q_j(t)\|$, should also in the the range of $(d,r)$ to avoid collision and ensure connectivity.
This problem is a special case of the classical capacitated facility location problem (CFLP) which is known to be NP-hard \cite{nphard}. Therefore, the problem \textbf{P0} is also an NP-hard problem and we need to resort to heuristic approaches to obtain acceptable solutions.

%\subsection{Problem Relaxation under Practical Scenarios}

%As the 3D coordinate space and velocity space are infinite and running time is continuous, the set of trajectories $\left<q_i(t),p_i(t)\right>, i\in\mathcal{U}$ is an exponentially large solution space. However, in practical circumstances, we assume the following relaxations to the problem \textbf{P0}:
%\begin{enumerate}
%    \item The running time is discretized with a minimal time step $\Delta t$ and UAVs have a maximum running time $t_{\max}$ limited by the battery capacity. This means $t=k\Delta t,k\in Z^+$ and $0\leq t \leq t_{\max}$. 
%    \item The maximum velocities of the UAVs are limited, which means $\|p_i\|\leq p_{\max}, \forall i\in \mathcal{U}.$
%\end{enumerate}
%With the above relaxations, the coordinate space, velocity space and time space are bounded in this problem.

\vspace{-0.0in}
\section{Methodology} \label{Sec:Methodology}

In this section, we use a dynamic and distributed approach to solve the optimization problem described by \textbf{P0}. The UAV network is orchestrated in real-time according to the QoS needs of the users through a controller-based UAV swarming approach, which autonomously creates a resilient configuration and leverages channel switching to provide customized service to both regular and premium users. We use the widely accepted kinematic model in robotics and control literature to describe the dynamics of the UAVs as $ \dot{q_i}=p_i$, and $\dot{p_i}=u_i$, where $u_i$ is the control input, $q_i,p_i,u_i \in \mathbb{R}^3$ and $i \in \mathcal{U}$. For the UAVs to organize themselves in a distributed manner, the control input $u_i$ is designed as a linear combination of the following three terms:
\vspace{-0.0in}
\begin{equation}\label{eq:4}
    u_i=f_i+g_i+h_i,
\end{equation}
where $f_i$ is an inter-UAV attractive/repulsive term, $g_i$ is a velocity consensus term, $h_i$ is a term defining the individual goal of each UAV. In order to create smooth incentive functions with finite cut-offs, we leverage the bump function $\alpha_{\{\gamma\}}(z)$ in \cite{bump_func} and $\sigma$-norm in \cite{multi_layer}. The bump function is defined as follows:
\begin{multline}
    \alpha_{\{\gamma\}}(z)=
    \left\{  
        \begin{array}{ll}  
        1,& \text{if}\ 0\leq z<\gamma,\\  
        \frac{1}{2}\left(1+\cos(\pi\frac{z-\gamma}{1-\gamma})\right),& \text{if} \ \gamma\leq z< 1,\\  
        0,& \text{if} \ z\geq 1,   
        \end{array}  
    \right.
\end{multline}
where $\gamma\in[0,1]$ and the $\sigma$-norm is defined as $\|z\|_\sigma=\frac{1}{\epsilon} \left( \sqrt{1+\epsilon\|z\|^2}-1 \right)$, where $\epsilon>0$ is a positive constant. The gradient of the $\sigma$-norm can be expressed as $\nabla \|z\|_\sigma=\frac{z}{\sqrt{1+\epsilon \|z\|^2}}=\frac{z}{1+\epsilon\|z\|_\sigma}$. We make use of these function to create incentives for the UAVs to achieve the optimal solution to \textbf{P0}. We describe each of these incentive functions along with their role in the following subsections.

\subsection{UAV Swarming}
The swarming behavior requires UAVs to maintain a certain distance with each other and follow a group objective. This is achieved by introducing the attraction/repulsion and velocity consensus functions, which maintains the distance between UAVs through the control of their velocity. Considering the objective to enhance QoS for users, a UAV $i$ should have the tendency to approach another UAV $j$ if UAV $j$ exceeds its serving capacity. Thus, the attractive/repulsive function can be written as follows \cite{resilient_connectivity_tccn,cognitive_connectivity}:
\vspace{-0.0in}
\begin{multline}
    f_i(\mathbf{q},\mathcal{M}^c)=\sum_{j\in N_i}\Bigg[\Phi (\|q_j-q_i\|_\sigma)+\\ a\left(1-\alpha_{\{0\}}\left(\frac{\|(|\mathcal{M}^c_i|-N^{\max})^+\|_\sigma}{\|N^{\max}\|_\sigma}\right)\right) \Bigg]\textbf{v}_{i,j},
\end{multline}
where $\textbf{v}_{i,j}=\nabla \|q_j-q_i\|_\sigma$ is the vector from $q_i$ to $q_j$. The function $\Phi(z)$ is expressed as:
\begin{equation}
    \Phi(z)=\alpha_{\{0.2\}}\left(\frac{z}{\|r\|_\sigma}\right)\phi(z-\|d\|_\sigma),
\end{equation}

\vspace{0.1in}where $\phi (z)=\frac{1}{2}[(a+b)\frac{(z+c)}{\sqrt{1+(z+c)^2}}+(a-b)]$ and $c=|a-b|/\sqrt{4ab}$ to ensure that $\phi (0)=0$. Here, $r$ is the maximum communication range and $d$ is the minimum distance between UAVs. The velocity consensus function works as a damping force that leads to a match of movements between neighboring UAVs. The function $g_i(\mathbf{p},\mathcal{M}^c)$ is in the form:
     
    \vspace{-0.1 in}
    \begin{equation}
        g_i(\mathbf{p},\mathcal{M}^c)=\sum_{j\in N_i}
        \alpha_{\{0.2\}}\left(\frac{\|q_i-q_j\|_\sigma}{\|r\|_\sigma}\right)(p_j-p_i).
    \end{equation}
    
    This enforces the alignment between UAVs to increase as the distance between them decreases, until reaching the maximum value at a distance of $0.2r$. The velocity consensus function reduces potential collisions and disconnections between UAVs during real-time operations.

% \begin{figure}[t]
%     \centering
%     \includegraphics[trim={12 0 10 0 in},clip,width=2.5in]{forces.pdf}
%     \caption{Illustration of forces among UAVs and users. The magnitudes of forces are proportional to their sizes in the figure. Premium users can achieve higher data rates by attracting the UAVs towards them. \vspace{-0.1in}}
%     \label{fig:forces}
% \end{figure}

\begin{algorithm}[h]
	\caption{QoS-Driven Dynamic Orchestration}
	\label{alg:UAV_control}
	\begin{algorithmic}[1]
	\Require Initialize position, velocity, and channel for each UAV $q_i(0)$, $p_i(0)$, $\mathcal{L}_i(0)\gets L_0$; Assign user category $\chi(m)$ and target data rate $\rho^{\chi(m)}$, record mean data rates $\mu_m$ in last $\tau$ seconds for each user $m\in \mathcal{M}$.
	\While{target data rates not satisfied}
	\For{each UAV $i$}
	\State{Determine the set of connected users $\mathcal{M}_i^C$.}
	\For{user $m\in\mathcal{M}_i^C$}
	\State{Compute the SINR and data rates $\mathcal{C}_m$ for each \\ \hspace{0.63 in}user $m$ using~\eqref{eq:sinr} and~\eqref{eq:data rates}.}
	    \If {$\chi= (\text{prem})$  \textbf{\&} $\mathcal{C}_m<\rho^{\chi(m)}$ \textbf{\&} $\mathcal{C}_m\leq \mu_m$}
	        \State{Premium user $m$ and UAV $i$ select a new\\ \hspace{0.84 in}available channel for communication.\nonumber} 
	    \EndIf
	    \State Recompute mean data rates $\mu_m$.
	\EndFor
	\State Compute control input $u_i$ for UAV $i$ using~\eqref{eq:4}.
	\State{Update the position $q_i$ and velocity $p_i$ using the\\ \hspace{0.42 in}discretized UAV dynamics}.
	\EndFor
    \EndWhile
	\end{algorithmic} 
\end{algorithm}
\vspace{-0.05 in}
\subsection{QoS-Driven Dynamic Orchestration}

The QoS requirements of users may vary by their subscribed services and application scenarios. We define a QoS-driven incentive function which aims to improve the received data rates of users. Based on target data rates $\rho^{\chi(m)}$ and achieved data rates $\mathcal{C}_m$ for each user $m\in\mathcal{M}_i^c$ connected to UAV $i$, this term reflects the affinity of a UAV and its connected users: \vspace{-0.0in}
\begin{equation}
\begin{aligned}
    h_i(\mathbf{q},\mathcal{M}^c)=
    \sum_{m\in\mathcal{M}\backslash\mathcal{M}_i^c}c_1\left(\frac{\rho^{\chi(m)}-\mathcal{C}_{m}}{\rho^{\chi(m)}}\right)^+\nabla\|q_i-q_m\|_\sigma \\
    +\sum_{m\in \mathcal{M}_i^c}c_2^{(\chi)}\alpha_{\{0\}}\left(\frac{\mathcal{C}_m}{\beta\rho^{\chi(m)}}\right)\phi(\rho^{\chi(m)}-\mathcal{C}_m)\nabla\|q_m-q_i\|_\sigma.
\end{aligned}
\label{eq:15}
\end{equation}

The first part of the equation represents a repulsive force from users against other UAVs. If a user's data rate requirement is not satisfied, it may generate a repulsive force against UAVs that it is not connected to for reducing interference. The second part of the equation represents an attractive force from users that are connected with UAV $i$ towards UAV $i$. If a user's data rate requirement is not satisfied, it may generate an attractive force to the UAV it connects with to improve link quality. Note that the parameter $c_2$ is selected differently for regular and premium users, i.e., $c_2^{(\text{prem})}=1.5c_2^{(\text{reg})}$. %The illustration of the forces on UAVs can be shown in Fig. \ref{fig:forces}

The framework for QoS-driven dynamic placement of UAVs can be summarized using Algorithm \ref{alg:UAV_control}. We assume each user always connect with the nearest UAV from its current position. In every iteration, UAVs compute data rates for all connected users and keep records of them. If the target data rates for a premium user is not reached and the real-time data rates have not increased during the past $\tau$ seconds, the UAV will switch to a free channel to reduce interference from other UAVs and therefore increase the data rates. The control input $u$ is computed using \eqref{eq:4} and is used to update position of the UAVs. The proposed distributed algorithm has linear time complexity of $\mathcal{O}(\|N\|+\|\mathcal{M}\|)$ for each UAV, where $N$ is the set of neighboring UAVs and $\mathcal{M}$ is the set of all users within the communication range of the UAV.

\vspace{-0.0in}
\section{Simulation Results} \label{Sec:Results}
In this section, we simulate different UAV network scenarios and validate the efficacy of our proposed algorithm. Unless otherwise specified, the model parameters used in the simulations are provided as follows: the height of all UAVs, $H = 100$ m
%\footnote{We restrict ourselves to horizontal placement of UAVs for ease of illustration of the experiments. However, the approach is directly applicable for 3D placement with bounds on the elevation.}
; the minimum distance between UAVs, $d=100$ m; the maximum communication range of UAVs, $r=300$ m. The target data rates for premium and regular users are set to $\rho^{(\text{prem})}=300$ Mbps and $\rho^{(\text{reg})}=100$ Mbps respectively. Other parameters and their values are tabulated in Table 1.

\begin{table}
\vspace{0.15 in}
\centering
\begin{tabular}{ |c|c|c| } 
 \hline
 \textbf{Parameter} & \textbf{Notation} & \textbf{Value} \\
 \hline
 UAV carrier frequency & $f_c$ & 2 GHz \\
 \hline
 LoS additional avg loss & $\eta_{LoS}$ & 0.1 dB \\ 
 \hline
 NLoS additional avg loss & $\eta_{NLoS}$ & 21.0 dB \\ 
 \hline
 Environment parameters & $\vartheta,\xi$ & 4.88, 0.43 \\
 \hline
 UAV transmission power & $P_t$ & 37.0 dBm \\
 \hline
 Bandwidth & $B$ & 15 MHz \\
 \hline
 Thermal Noise & $\sigma^2$ & $-$80 dBm \\
 \hline
 Incentive function & $\epsilon,a,b,N^{\max}$ & 0.1, 5.0, 5.0, 80.0 \\
 parameters & $c_1,c_2^{(\text{reg})},\beta$ & 6.0, 4.0, 1.5 \\
 \hline
 Channel switch interval & $\tau$ & 5.0 sec \\
 \hline
\end{tabular}
\caption{Simulation parameters} \vspace{-0.2 in}
\end{table}

\begin{figure}[t]
    \centering
    \vspace{0.0in}
    \includegraphics[width=\linewidth]{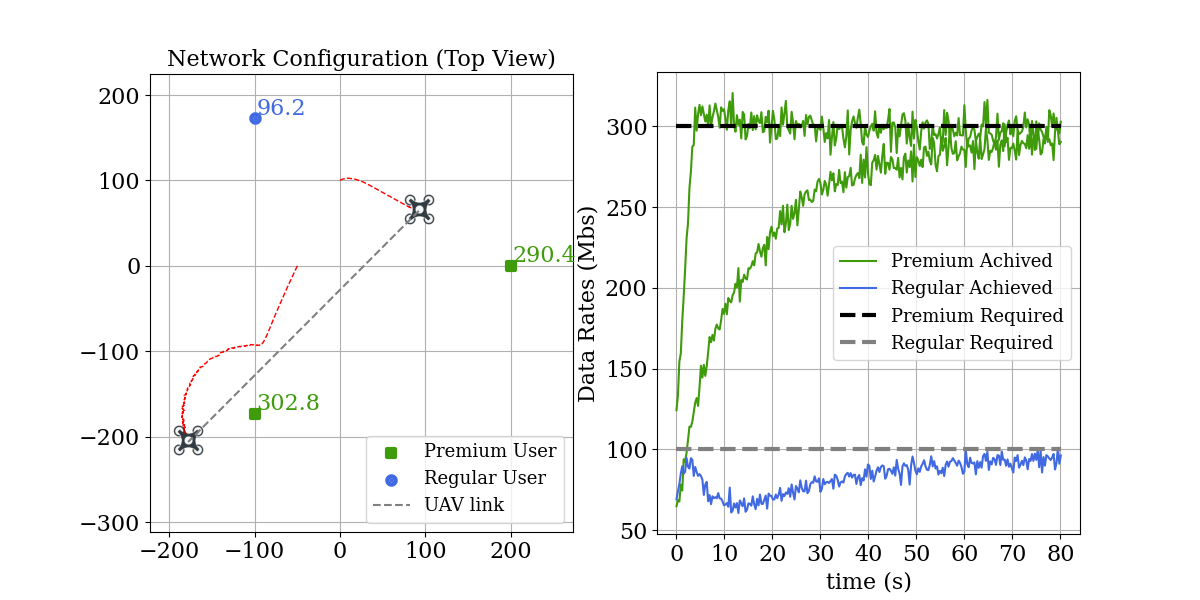}
    \caption{\vspace{-0.0in}Simulation results with 3 users (2 premium and 1 regular) and two UAVs using the proposed method. Red lines show the trajectories of UAVs. All three users achieve close to target data rates on convergence.}
    \label{fig:result1}
    \vspace{-0.2in}
\end{figure}
\begin{figure}[t]
    \centering
    \vspace{0.0in}
    \includegraphics[width=\linewidth]{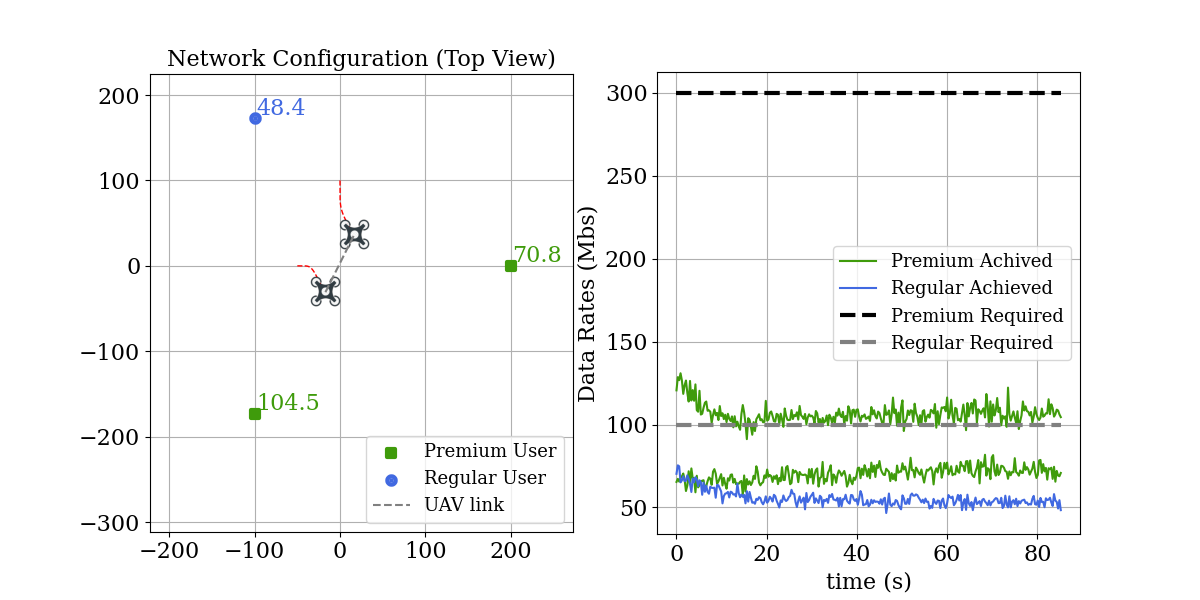}
    \caption{\vspace{-0.0in}Comparison simulation using flocking algorithm only. The users cannot achieve target data rates.}
    \label{fig:result1_compare}
    \vspace{-0.2in}
\end{figure}

\subsection{QoS Driven Proactive UAV Placement}
We first consider a simple scenario of three users served by two UAVs to demonstrate the operation of the proposed framework. The user locations and types are shown in Fig. \ref{fig:result1}. We consider two premium users and one regular user. UAVs are initially placed arbitrarily and the red dotted lines shows how their trajectory evolve over time to reach the final configuration. At the initial state, 
% the regular user achieves higher data rates than its target, so it generates a repulsive force against the UAV it is connected to. On the other hand, 
the premium users do not achieve the required data rates so they generate an attractive force on their connected UAVs. Hence, we observe that the two UAVs move closer to the respective premium users. However, to further enhance data rates, the UAVs need to move farther from each other to reduce interference. Therefore, one of the UAVs follows an arc trajectory around the connected user to maximize the data rates of the regular user while maintaining the already achieved data rates of the premium user. Note that the UAVs also need to ensure a minimum distance between then to enable localized backhaul connectivity. This is highlighted by the black dotted line in the figure. Simulation using standard flocking algorithm \cite{multi_layer} alone was compared in Fig. \ref{fig:result1_compare}. The UAVs can gather around the centroid of the three users and maintain the minimum distance with each other. However, they cannot adapt their positions according to the QoS needs and resulted in unsatisfied data rates.

% \begin{figure}[t]
%     \centering
%     \vspace{-0.0in}
%     \includegraphics[trim={60 5 80 20 in},clip,width=\linewidth]{Figure_4.pdf}
%     \caption{Simulation results with 1 UAV, 1 premium user, and 1 regular user. The regular user achieves its target data rate while the premium user achieves $\sim$97\% of its target data rate.\vspace{-0.2in}}
%     \label{fig:result1}
% \end{figure}

% Fig. \ref{fig:result2}, \ref{fig:result3} and \ref{fig:result4} show the experimental results for 3 users and 4 users. The target data rates for premium users and regular users are set to 8 Mbps and 4 Mbps respectively. In Fig. \ref{fig:result2} and \ref{fig:result3}, it can be observed that all users' QoS requirements are achieved upon convergence. In Fig. \ref{fig:result4}, two premium users receive over 92\% of their target QoS while QoS requirements for the others are completely achieved.

\subsection{Resilience under Failures and Attacks}
\begin{figure}[t]
    \centering
    \vspace{0.0in}
    \includegraphics[width=0.99\linewidth]{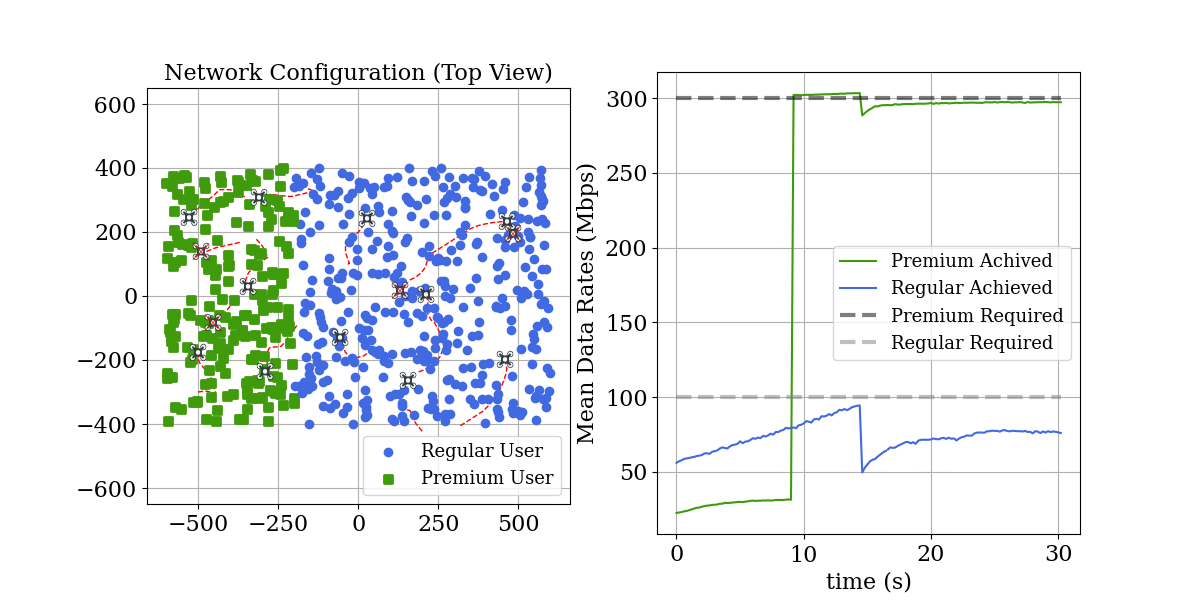}
    \caption{Simulation results with 600 users randomly distributed in a rectangular space. 30\% of users are premium users and the rest are regular users. 30\% of UAVs which randomly failed at t=15s were labeled with red cross.\vspace{-0.0in}}
    \label{fig:result2}
    \vspace{-0.2in}
\end{figure}

In Fig. \ref{fig:result2}, we consider the scenario of 600 users located in a rectangular region mimicking a parade or a procession. The users are placed randomly according to a uniform distribution and are split into two types. We assume that 30\% of the users are premium users and are located in the left part highlighted by green dots. The regular users are located on the right side and highlighted in blue. 15 UAVs are deployed to provide service to the users. The initial positions and traces of the UAVs are shown by dotted red lines in Fig. \ref{fig:result2}. Due to the signal interference and geographical distance, the premium users gain low data rates in the initial state. At time $t=9.0$s, the data rates of premium users reach a bottleneck so UAVs which serve premium users switch their channels because the premium users cannot meet their QoS requirements. Seven UAVs change to serve premium users only and the other three UAVs serve the regular users. As the interference is eliminated, the mean data rates for premium users goes up to the desired 300 Mbps at time $t=10.0$s. To test the resilience of the network, 30\% of UAVs randomly fail at t=15s. Failed UAVs are labeled with red cross. The UAVs rapidly reconfigure their formation and recover the QoS. At t=30s, the mean data rates for premium users recover to over 293 Mbps and the data rates for regular users recover to over 79 Mbps. 

\subsection{Network Size and Impact on QoS Provisioning}
\begin{figure*}[t]
\centering
\begin{subfigure}{0.32\linewidth}
    \centering
    \vspace{-0.0in}
    \includegraphics[width=\linewidth]{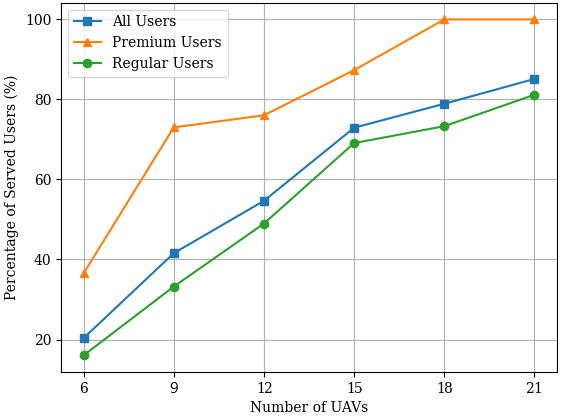}
    \caption{}
    %\caption{Percentage of served users for different user groups against the number of UAVs.\vspace{-0.1in}}
    \label{fig:result3}
\end{subfigure}
\begin{subfigure}{0.32\linewidth}
    \centering
    \vspace{-0.0in}
    \includegraphics[width=\linewidth]{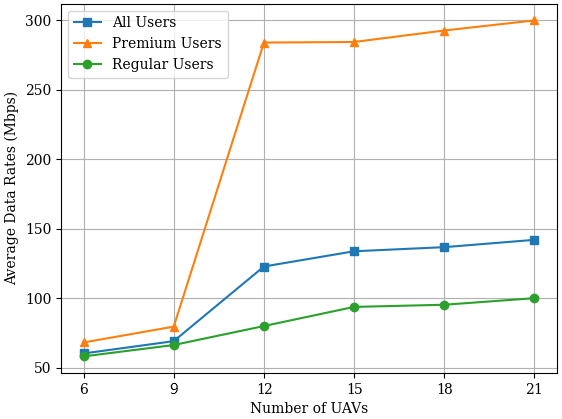}
    \caption{}
    %\caption{Average data rates for different user groups against the number of UAVs.\vspace{-0.1in}}
    \label{fig:result4}
\end{subfigure}
\begin{subfigure}{0.32\linewidth}
    \centering
    \vspace{-0.0in}
    \includegraphics[width=\linewidth]{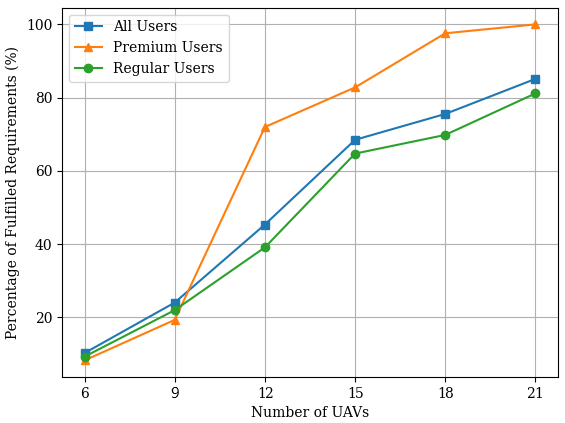}
    \caption{}
    %\caption{Percentage of fulfilled data rate requirements for different user groups against the number of UAVs.\vspace{-0.1in}}
    \label{fig:result5}
\end{subfigure}
\caption{Percentage of served users, average data rates and percentage of fulfilled data rate requirements for different user groups against the number of UAVs.\vspace{-0.2in}}
\end{figure*}

Next, we investigate the impact of changing the number of UAV BSs on the the achieved performance of the ground users. We use the scenario illustrated in Fig.~\ref{fig:result2} and increase the number of UAVs to observe the number of served users and achieved data rates as compared to the targets of both premium and regular users. %we compared the QoS of the network using different numbers of UAVs serving the ground users. 
Fig. \ref{fig:result3}, \ref{fig:result4}, and \ref{fig:result5} respectively show the percentage of served users, average data rates, and percentage of fulfilled data requirements for different user groups against the number of UAVs. When only 6 UAVs are available, the UAVs can only serve a small proportion of the total users due to capacity constraints and can be far away from users that are located at the edges. So the percentage of served users and the average data rates are low. As the number of UAVs increases to 21, the QoS of the network improves until all the premium users are served and achieve the required data rates. It can be shown that the QoS of the network makes fast improvements at initial states before converging to their maximum values as the number of UAVs increases until reaching the capacity limit of the UAV network. Note that the premium users achieve higher performance not only in terms of the target data rates but also in the percentage of served users and the percentage of fulfilled requirements. The only exception to this is observed in Fig.~\ref{fig:result5} for small number of UAVs since the target data rates of premium users are too aggressive and could not be achieved by a small number of UAVs. However, with large number of UAVs, both the primary users and regular users are able to completely achieve their targets.

\vspace{-0.0in}
\section{Conclusions} \label{Sec:Conclusion}
In this paper, we investigated the design of QoS-centric dynamic placement of UAVs to provide customized service quality for different user requirements. We leveraged the mobility and flexibility of aerial placement in UAVs and developed on-demand UAV small cell deployment strategies to not only provide higher data rates to users in required regions but also provide service differentiation for meeting users needs (live streaming, social media activities) and also higher revenues for service providers. The proposed method utilized distributed UAV flocking and swarming dynamics, and extended it by adding an incentive function based on target and real-time data rates. Compared with open-loop optimization approaches, the QoS-driven dynamic placement algorithm has linear time complexity and provides adaptive coverage and differentiated service levels. The simulation results show that our proposed method is able to fulfill the data rates requirements of premium and regular users in both simple and complex scenarios. The proposed solution can be useful in real networks where fixed small cell deployment may be economically in-viable for occasional needs for high speed connectivity. 

\ifCLASSOPTIONcaptionsoff
  \newpage
\fi

\vspace{-0.0in}
\bibliographystyle{IEEEtran}
\bibliography{IEEEabrv,Bibliography}

\end{document}